\documentstyle[epsfig,12pt]{article}

\begin{document}
\begin{center}

CORRELATOR ANALYSIS OF MULTIPARTICLE EVENTS

\vskip 3mm

N. Amelin, P. Filip, R. Lednicky, M. Pachr

\vskip 3mm

{\small
 {\it
 Joint Institute for Nuclear Research, \\
Dubna, Moscow Region, 141980, Russia}
}


\end{center}

\vskip 3mm

\begin{center}

\begin{abstract}

A procedure for the evaluation of correlators
of any order in a reasonable computer time is presented.
Connection between correlators and fluctuations of the
event mean values of observables is discussed.
Extension of the procedure to event-by-event
approach is suggested. The usefulness of the method
is demonstrated using the events
simulated within various models of multipaticle production.

\end{abstract}

\vskip 5mm

{\bf Keywords:}
particle correlations, correlators, mean value fluctuations
\end{center}

\vskip 10mm

\section{Introduction}

Recent and near-future experiments devoted to the study of
relativistic nuclear collisions (see, e.g.,
\cite{Voloshin2002,STAR2003})
have stirred up
considerable interest in suitable methods of the analysis of
high particle multiplicity  events.
For example, fluctuations of various observables are widely
studied using the event-by-event approach \cite{QM04}.
In this paper we consider the integral characteristics of
particle correlations,
namely the correlators \cite{MS2001} for a given
particle observable (e.g., the particle energy $E$,
transverse momentum $p_t$ or rapidity $y$).
We suggest a fast procedure how to construct these quantities.
We also investigate their properties,
including the connection with the
event-to-event fluctuations, and their ability to reflect
underlying mechanisms in the case of several models.

The paper is now organized as follows.
In section 2 we give the formal definition of correlators
and discuss some of their properties.
Sections 3-5 deal with the evaluation of correlators from
experimental data.
In section 4 we also consider the connection of the
correlators with the fluctuations of the observable event-mean values.
In section 6 we apply the proposed procedure to events
simulated within various models of particle production.
The results are summarized in section 6.
In Appendix we derive the expressions allowing one to relate the
correlators with
the central moments of the single-particle distributions and the
event-to-event fluctuations of the observable mean.

\section{Particle correlators}

Let us start with events of fixed multiplicity
$n$ of the produced or observed particles of a given kind.
The inclusive or exclusive production probability
of $\nu \leq n$ identical particles can be described by a
multivariate probability distribution function (PDF)
$f_\nu^{(n)}(x_1,\dots,x_\nu)$ of a given observable $x$,
where $f_\nu^{(n)}$ is a symmetric
function normalized to unity:

\begin{equation}
\label{normPDF}
\int f_\nu^{(n)}(x_1,\dots,x_\nu)dx_1\dots dx_\nu = 1.
\end{equation}
The PDF can be characterized by the mean value
\begin{equation}
\label{mean} \bar{x}^{(n)}=\int x f_1^{(n)}(x)dx,
\end{equation}
and - by the $l$-order moment-type quantities
\begin{equation}
\label{C3} C^{(n,x)}_{i_1,\dots,i_l} =
\int (x_{i_1} - \bar{x}^{(n)})\dots
(x_{i_l}-\bar{x}^{(n)})f_n^{(n)}(x_1,\dots,x_n)dx_1\dots dx_n.
\end{equation}
Due to the PDF symmetry, all the quantities
$C^{(n,x)}_{i_1,\dots,i_l}$
coincide for any permutation of indexes.

In the case of independent production of particles
of a given kind the multivariate PDF reduces to the
product of the same one-particle distribution functions
(neglecting the correlations due to quantum statistics and
final state interaction):
\begin{equation}
\label{f} f_\nu^{(n)}(x_1,\dots,x_\nu)=
f_1^{(n)}(x_1)\dots f_1^{(n)}(x_\nu).
\end{equation}
Obviously, in this case
$C^{(n,x)}_{i_1,i_2,...i_l}=0$ for $i_1\neq i_2\dots\neq i_l$.
The quantities $C^{(n,x)}_{i_1,i_2,\dots,i_l}$ for
$i_1\neq i_2\dots\neq i_l$ thus measure a correlation
among produced particles and
therefore we will call them correlators.
Since correlators do not depend on the particular set of
$l$ mutually different indexes, we will use for them the simple
notation $C^{(n,x)}_l$.
Note that for equal indexes
$i_1=i_2=\dots=i_l$,
$C^{(n,x)}_{i_1,i_2,...i_l}\equiv M^{(n,x)}_l$
represents the $l$-th central moment of the one-particle PDF.

In practice, number $n$ of produced or observed particles
of a given kind can vary from 0 to the ultimate multiplicity
$n_{\rm max}$,
so one has to introduce the corresponding probabilities
$p_n$ and rewrite Eqs.~(\ref{mean}) and (\ref{C3}) as
\begin{equation}
\label{mean'}
\langle \bar{x}\rangle\equiv
\left\langle \bar{x}^{(n)}\right\rangle=
\sum_{n=1}^{n_{\rm max}} p_n \bar{x}^{(n)}/
\sum_{n=1}^{n_{\rm max}} p_n=
\sum_{n=1}^{n_{\rm max}} p_n
\int x f_1^{(n)}(x)dx/
\sum_{n=1}^{n_{\rm max}} p_n,
\end{equation}
$$ C^{(x)}_{l} =\sum_{n=l}^{n_{\rm max}} p_n\cdot$$
\begin{equation}
\label{C3'}
\cdot\int (x_{1} - \langle \bar{x}\rangle)\dots
(x_{l}-\langle \bar{x}\rangle )f_l^{(n)}(x_1,\dots,x_l)
dx_1\dots dx_l
/\sum_{n=l}^{n_{\rm max}} p_n.
\end{equation}

Generally, the PDF can depend on various event characteristics
$\alpha$, including the observed multiplicity, the particle
composition or the selected range of the impact parameters.
Eqs.~(\ref{mean'}) and (\ref{C3'}) should be then generalized
by the substitution $\sum p_n \to \sum p_\alpha$.
Particularly,
\begin{equation}
\label{mean'a}
\langle \bar{x}\rangle =
\left\langle \bar{x}^{(n)}\right\rangle=
\sum_{\alpha} p_\alpha \bar{x}^{(\alpha)}/
\sum_{\alpha} p_\alpha=
\sum_{\alpha} p_\alpha \int x f_1^{(\alpha)}(x)dx/
\sum_{\alpha} p_\alpha .
\end{equation}

Using the identity
\begin{equation}
\label{C25mpa}
x_{k}-\langle \bar{x}\rangle =
(x_{k}-\bar{x}^{(\alpha)})+
(\bar{x}^{(\alpha)}-\langle \bar{x}\rangle),
\end{equation}
one can rewrite the global correlator $C_l^{(x)}$
in terms of the $\alpha$-dependent correlators
\begin{equation}
\label{C3'a}
C^{(\alpha,x)}_{l} =
\int (x_{1} - \langle \bar{x}^{(\alpha)}\rangle)\dots
(x_{l}-\langle \bar{x}^{(\alpha)}\rangle )
f_l^{(\alpha)}(x_1,\dots,x_l)
dx_1\dots dx_l
\end{equation}
and the fluctuations of the observable mean $\bar{x}^{(\alpha)}$
at a given $\alpha$ around the global mean
$\langle \bar{x}\rangle$:
$$
C^{(x)}_l =\sum_\alpha p_\alpha
\sum_{\lambda=0}^{l} {l\choose \lambda}C^{(\alpha,x)}_{\lambda}
(\bar{x}^{(\alpha)}-\langle \bar{x}\rangle)^{l-\lambda}/
\sum_\alpha p_\alpha
$$
\begin{equation}
\label{C25''a}
\equiv\left\langle
\sum_{\lambda=0}^{l} {l\choose \lambda}C^{(\alpha,x)}_{\lambda}
(\bar{x}^{(\alpha)}-\langle \bar{x}\rangle)^{l-\lambda}
\right\rangle,
\end{equation}
where $C^{(\alpha,x)}_0=1, C^{(\alpha,x)}_1=0$.
One may see that the absence of the correlation at any
$\alpha$ (i.e. $C^{(\alpha,x)}_{l} =0$ for $l>1$) does not lead to a
vanishing global correlator. In this case, the latter is
solely determined by the fluctuation of the $\alpha$-dependent
observable mean: $C^{(x)}_l =
\langle (\bar{x}^{(\alpha)}-\langle \bar{x}\rangle)^l \rangle$.

The above formalism ignores the possible non-identity
of the selected particles. Its generalization to the correlators
of different particle species is however straightforward.
For example, for two types of particles, say those characterized
by positive ($+$) and negative ($-$) charge, one has to make the
substitutions $l\to l_+,l_-$,
$x\to x_+$ and $x_-$. Particularly,
\begin{equation}
\label{mean+-}
\bar{x}_+^{(\alpha)}=\int x_+ f_{1,0}^{(\alpha)}(x_+)dx_+,~~~
\bar{x}_-^{(\alpha)}=\int x_- f_{0,1}^{(\alpha)}(x_-)dx_-,
\end{equation}
$$
C^{(\alpha,x)}_{l_+,l_-} =
\int (x_{+_{1}} - \langle \bar{x}_+^{(\alpha)}\rangle)\dots
(x_{+_{l_+}}-\langle \bar{x}_+^{(\alpha)}\rangle )
(x_{-_{1}} - \langle \bar{x}_-^{(\alpha)}\rangle)\dots
(x_{-_{l_-}}-\langle \bar{x}_-^{(\alpha)}\rangle )\cdot
$$
\begin{equation}
\label{C3'a+-}
f_{l_+,l_-}^{(\alpha)}(x_{+_{1}},\dots,x_{+_{l_+}};x_{-_{1}},\dots,x_{-_{l_-}})
dx_{+_{1}}\dots dx_{+_{l_+}}dx_{-_{1}}\dots dx_{-_{l_-}}.
\end{equation}

It is instructive to express the correlator
$C_2^{(x)}\equiv C_{cc}^{(x)}$ of two
charged particles in the events with $n=n_+ + n_-$ selected
particles through the correlators $C_{2,0}^{(x)}\equiv C_{++}^{(x)}$,
$C_{0,2}^{(x)}\equiv C_{--}^{(x)}$ and
$C_{1,1}^{(x)}\equiv C_{+-}^{(x)}$. Using the identity
\begin{equation}
\label{ident+-}
\bar{x}=\bar{x}_\pm - (\bar{x}_\pm -\bar{x})
\equiv\bar{x}_\pm - \Delta\bar{x}_\pm ,
\end{equation}
one can write
\begin{equation}
\label{Ccc+-}
C_{cc}^{(x)}=\frac{n_{++}}{n_{cc}}[C_{++}^{(x)}+\Delta\bar{x}_+{}^2]+
\frac{n_{--}}{n_{cc}}[C_{--}^{(x)}+\Delta\bar{x}_-{}^2]+
\frac{n_{+-}}{n_{cc}}[C_{+-}^{(x)}+\Delta\bar{x}_+\Delta\bar{x}_-],
\end{equation}
where
\begin{equation}
\label{ncc}
n_{cc}=n_{++}+n_{--}+n_{+-}=n_+(n_+ -1)/2+n_-(n_- -1)/2+n_+n_-
\end{equation}
is the number of charged pairs.
One may see that even in the absence of correlations of particles
of given charges, i.e. $C_{++}^{(x)}=C_{--}^{(x)}=C_{+-}^{(x)}=0$,
the correlator $C_{cc}^{(x)}$ can be non-zero provided
$\bar{x}_+\ne \bar{x}_-$.
In particular, for $n_+=n_-=n/2$,
\begin{equation}
\label{mean_cc}
\bar{x}=\frac{n_+}{n}\bar{x}_+ +\frac{n_-}{n}\bar{x}_- =
\frac12(\bar{x}_+ +\bar{x}_-),~~~
\Delta\bar{x}_+=-\Delta\bar{x}_- =\frac12(\bar{x}_+ -\bar{x}_-)
\end{equation}
and, in the absence of correlations,
\begin{equation}
\label{C0_cc}
C_{cc}^{(x)}=-\frac{(\bar{x}_+ -\bar{x}_-)^2}{4(n-1)} \le 0.
\end{equation}

\section{Correlator estimates}

Let us first consider $n_{\rm evt}$ experimental events with
a fixed multiplicity $n$.
The mean value of the observable $x$ can then be
estimated as

\begin{equation}
\label{C4} \langle \bar{x}\rangle  =
\frac{1}{n_{\rm evt}}\sum_{i = 1}^{n_{\rm evt}}
\bar{x}^{(i)},
\end{equation}
where $\bar{x}^{(i)}$ is the estimate of the observable mean
in the $i$-th event:
\begin{equation}
\label{C4'} \bar{x}^{(i)} =\frac{1}{n}\sum_{j = 1}^{n}x^{(i)}_j.
\end{equation}
Similarly, the correlator $C^{(x)}_l$ can be estimated as
\begin{equation}
\label{C25} C^{(x)}_l = \frac{1}{n_{\rm evt}}
\sum_{i = 1}^{n_{\rm evt}}
C^{(i,x)}_l,
\end{equation}
where $C^{(i,x)}_l$ is the estimate of the correlator in the
$i$-th event:
\begin{equation}
\label{C25'} C^{(i,x)}_l =
\frac{1}{n_l}\sum(x^{(i)}_{i_1} - \langle \bar{x}\rangle )
\dots (x^{(i)}_{i_l} - \langle \bar{x}\rangle );
\end{equation}
the sum in Eq.~(\ref{C25'}) runs over all $n_l$ sets of
$l$ particles
chosen from $n$ particles in the event.
One can sum either over $n_l={n\choose l}=n(n-1)\dots (n-l+1)/l!$
of the sorted sets $i_1<i_2 <\dots <i_l$
or over $n_l=n(n-1)\dots (n-l+1)$ of the
unsorted sets $i_1\ne i_2\ne\dots\ne i_l$. In the latter case,
each of the unsorted sets gives rise to the $l!$ of identical
terms in the sum.

In the case of a mixture of events with different multiplicities
$n^{(i)}$, one can estimate the observable mean and the
correlators in a similar way as in the two-step averaging
procedure given in Eqs.~(\ref{C4})-(\ref{C25'}).
One should only make substitutions $n\to n^{(i)}$,
$n_l\to n^{(i)}_l$ in Eqs.~(\ref{C4'}), (\ref{C25'}) and
take into account that the single-event averages
enter into Eqs.~(\ref{C4}) and (\ref{C25}) multiplied by the
weights proportional to $n^{(i)}$ and $n^{(i)}_l$, respectively:
\begin{equation}
\label{C4a}
\langle \bar{x}\rangle
\equiv \left\langle \bar{x}^{(i)}\right\rangle
= \frac{1}{N}\sum_{i = 1}^{n_{\rm evt}}
n^{(i)}\bar{x}^{(i)},~~~N=\sum_{i=1}^{n_{\rm evt}} n^{(i)},
\end{equation}
\begin{equation}
\label{C25a} C^{(x)}_l
\equiv \left\langle C^{(i,x)}_l\right\rangle_l
= \frac{1}{N_l}
\sum_{i = 1}^{n_{\rm evt}} n^{(i)}_l
C^{(i,x)}_l,~~~N_l=\sum_{i=1}^{n_{\rm evt}} n^{(i)}_l.
\end{equation}
The same result can be obtained by averaging simply over all
$N$ collected particles or all $N_l$
$l$-particle sets (formed from particles within the same event
only):
\begin{equation}
\label{C26} \langle \bar{x}\rangle  = \frac{1}{N}
\sum_{j = 1}^N x_j,
\end{equation}
\begin{equation}
\label{C27} C^{(x)}_l = \frac{1}{N_l}
\sum(x_{i_1} - \langle \bar{x}\rangle )
\dots (x_{i_l} - \langle \bar{x}\rangle ),
\end{equation}
where the sum in Eq.~(\ref{C27}) runs over all $N_l$ sets.

The generalization of Eqs. (\ref{C4})-(\ref{C27}) to two or more
particle species is straightforward. For example, for two types
of particles characterized by positive and negative charge, one
has to make substitutions $l\to l_+,l_-$, $n\to n_+,n_-$,
$n_l\to n_{l_+},n_{l-}$, $N_l\to \sum_i n^{(i)}_{l_+}n^{(i)}_{l-}$
and $x\to x_+,x_-$. Particularly,
\begin{equation}
\label{C25'+-}
C^{(i,x)}_{1,1} =(\bar{x}_+^{(i)}-\langle x_+\rangle)
(\bar{x}_-^{(i)}-\langle x_-\rangle),~~~
C^{(i,x)}_{l_+,l_-} = C^{(i,x_+)}_{l_+}C^{(i,x_-)}_{l_-}.
\end{equation}

To calculate the errors, we can split all the events into $n_g$
subgroups, each with about the same number of events,
and estimate the correlator $C^{(m)}$
(we omit all other indexes for the sake of simplicity)
in the $m$-th subgroup using the global estimate of the
mean value $\langle x\rangle $.
The global estimate of the correlator is then given by the mean
of the group values:
\begin{equation}
\label{C29} C=\frac{1}{n_g}\sum_{m=1}^{n_g} C^{(m)}.
\end{equation}
The dispersion of the group values is
\begin{equation}
\label{C28} D =\frac{1}{n_g-1}\sum_{m = 1}^{n_g}
(C^{(m)}-{C})^2,
\end{equation}
where the sum of the deviations squared is divided by $n_g-1$
since one degree of freedom is used to determine the
global estimate of the correlator according to Eq.~(\ref{C29}).
Neglecting a small correlation of the group correlators due to
the use of the global estimate of $\langle x\rangle $,
one can calculate the error in the global correlator estimate as
\begin{equation}
\label{C30} \sigma({C})=\left(D/n_g\right)^{1/2}=
\left(\frac{1}{n_g(n_g-1)}\sum_{m=1}^{n_g}
(C^{(m)}-{C})^2\right)^{1/2}.
\end{equation}

\section{Correlators and event-to-event fluctuations}\label
{Sec_cor-fluct}

To relate the correlator $C_l^{(x)}$ with the event-to-event
fluctuation of the observable single-event mean $\bar{x}^{(i)}$,
one can use the analog of the identity in Eq.~(\ref{C25mpa})
with the substitutions $x_{k}\to x_{k}^{(i)}$ and
$\bar{x}^{(\alpha)}\to \bar{x}^{(i)}$,
and rewrite the estimate of the single-event correlator
in Eq.~(\ref{C25'}) in the form
\begin{equation}
\label{C25''} C^{(i,x)}_l =
\sum_{\lambda=0}^{l} {l\choose \lambda}c^{(i,x)}_{\lambda}
(\bar{x}^{(i)}-\langle \bar{x}\rangle)^{l-\lambda},
\end{equation}
where $c^{(i,x)}_0=1, c^{(i,x)}_1=0$ and $c^{(i,x)}_{l}$ for
$l\ge 2$ is defined similar to Eq.~(\ref{C25'})
except for the substitution
$\langle \bar{x}\rangle\to \bar{x}^{(i)}$:
\begin{equation}
\label{C25'''} c^{(i,x)}_l =
\frac{1}{n^{(i)}_l}
\sum(x^{(i)}_{i_1} - \bar{x}^{(i)})\dots
(x^{(i)}_{i_l} - \bar{x}^{(i)}).
\end{equation}
The estimate of the correlator
can then be written in the form
\begin{equation}
\label{C25''aa} C^{(x)}_l =
\left\langle
\sum_{\lambda=0}^{l} {l\choose \lambda}c^{(i,x)}_{\lambda}
(\bar{x}^{(i)}-\langle \bar{x}\rangle)^{l-\lambda}
\right\rangle_l ,
\end{equation}
where the $l$-dependent averaging is defined in
Eq.~(\ref{C25a}).
Introducing the notation
\begin{equation}
\label{diff}
\Delta\bar{x}^{(i)}=\bar{x}^{(i)}-\langle \bar{x}\rangle
\end{equation}
and omitting the event indexes, Eq.~(\ref{C25''aa})
particularly yields:
\begin{equation}
\label{cormn2}
C_2^{(x)}=\langle c_2^{(x)}  + \Delta\bar{x}^2\rangle_2,
\end{equation}
\begin{equation}
\label{cormn3}
C_3^{(x)}=\langle c_3^{(x)} +
3c_2^{(x)}\Delta\bar{x}  + \Delta\bar{x}^3\rangle_3,
\end{equation}
\begin{equation}
\label{cormn4}
C_4^{(x)}=
\langle c_4^{(x)} + 4c_3^{(x)}\Delta\bar{x}
+6c_2^{(x)}\Delta\bar{x}^2
+ \Delta\bar{x}^4\rangle_4.
\end{equation}
For two different particle species $+$ and $-$,
Eqs.~(\ref{C25'+-}) and (\ref{C25''aa}) yield, e.g.:
\begin{equation}
\label{cormn2+-}
C_{1,1}^{(x)}\equiv C_{+-}^{(x)}=
\langle\Delta\bar{x}_+\Delta\bar{x}_-\rangle_{1,1},~~~
C_{2,1}^{(x)}\equiv C_{++-}^{(x)}=
\langle (c_2^{(x_+)}  +
\Delta\bar{x}_+{}^2)\Delta\bar{x}_-\rangle_{2,1}.
\end{equation}

The meaning of quantities $c_l^{(x)}$ is clarified in
Appendix, where it is shown that they can be expressed
through the estimates of the moments of the single-particle
$x$-distribution in a given event:
\begin{equation}
\label{schc5} m_\lambda^{(x)} =
\frac{1}{n} \sum_{j=1}^n (x_j-\bar{x})^\lambda.
\end{equation}
Particularly,
\begin{equation}
\label{KK2}c_2^{(x)}
= - \frac{m_2^{(x)}}{n-1},
\end{equation}
\begin{equation}
\label{KK3}
c_3^{(x)}
= \frac{2m_3^{(x)}}{(n-1)(n-2)},
\end{equation}
\begin{equation}
\label{KK4}
c_4^{(x)}
= \frac{3 n \left(m_2^{(x)}\right)^2-6 m_4^{(x)}}
{(n-1)(n-2)(n-3)},
\end{equation}

One may conclude from Eqs.~(\ref{cormn2})-(\ref{cormn2+-}) and
Eqs.~(\ref{KK2})-(\ref{KK4}):
\begin{itemize}
\item
Quantities $c_l^{(x)}$
are determined by the shape of single-particle $x$-distribution
and by multiplicity of detected ($n$) or selected ($n\to \nu\le n$)
particles in a given event.
Therefore, they are sensitive only to a part of the
correlation related to this shape; an example is the correlation
due to energy-momentum conservation (see the 4-th item).
The remaining part is contained in the
event-to-event fluctuations of the observable mean in
accordance with Eqs.~(\ref{C25''aa})-(\ref{cormn2+-}).
\item
The magnitude of the quantities $c_l^{(\nu,x)}$ decreases with
the increasing number $\nu$ of selected particles. This decrease
should be compensated by the $\nu$-dependence of the fluctuations
of single-event mean observable $\bar x^{(i)}$
to guarantee the $\nu$-independence of the correlators.
Particularly, for uncorrelated production ($C_l^{(x)}=0$),
one gets
$\langle\Delta\bar{x}^2\rangle_2^{(\nu)}=\langle
m_2^{(x)}\rangle_2/(\nu-1)$.
\item
Assuming the moments $m_l^{(x)}$ of the single-particle
distribution weakly varying with the number $n$ of observed
particles, the correlators for high enough $n$ are dominated by
the event-to-event fluctuations, i.e.
$C_l^{(x)}\to\langle\Delta\bar{x}^l\rangle_l$.
\item
For a conserved additive observable $x$ (e.g., particle energy),
considering particles of any kind and assuming that all
$n_{\rm tot}$ particles are observed ($n=n_{\rm tot}$),
the mean $\bar{x}$ does not fluctuate so
$C_l^{(x)}=\langle c_l^{(x)}\rangle_l$.
Particularly, $C_2^{(x)}=-\langle
m_2^{(x)}\rangle_2/(n_{\rm tot}-1)$ and
$C_3^{(x)}=2\langle
m_3^{(x)}\rangle_3/[(n_{\rm tot}-1)(n_{\rm tot}-2)]$
for the events with about the same numbers of produced
particles.
\item
Evaluation of correlators $C_l^{(x)}$ using
Eqs.~(\ref{KK2})-(\ref{KK4})
can save substantial amount of computing time compared with
the direct evaluation according
to Eq.~(\ref{C27}).
In the former case
the number of operations is $\propto n$ while
in the latter case it is $\propto n^l$.
\end{itemize}

\section{Event-by-event correlators}
In the case of uncorrelated particle production,
the event-to-event fluctuations of the observable
mean are solely determined by the quantities $c_l^{(x)}$.
Such ``standard'' fluctuations are thus related
to the single-particle distributions
and can be estimated by the event-mixing techniques.
The eventual deviation from the ``standard''
or ``statistical'' fluctuations
then signals the presence of correlations or - a mixture
of the events with different single-particle distributions.

Therefore, to clarify the origin of the ``non-standard''
or ``non-statistical''
fluctuations, it is desirable to estimate the correlators
for the events with similar characteristics $\alpha$.
The ultimate solution is to use
the information from a given event only.
To do so, one has to destroy the equality in Eq.~(\ref{ident1}),
i.e. decouple as much as possible the observables $x_j$
entering into the correlator estimate from the observable
mean $\bar{x}$ in a given event. In the case of sufficiently
large multiplicity, it can be achieved by splitting
the event in a number of sub-events $s=1,2,\dots,n_{\rm sevt}$
with about the same multiplicities $n^{(s)}$. One can then
estimate the correlators according to
Eqs.~(\ref{C4})-(\ref{C27})
or (\ref{cormn2})-(\ref{cormn2+-}) and
(\ref{schc5})-(\ref{KK4}),
making the substitutions $i\to s$ and
$n_{\rm evt}\to n_{\rm sevt}$.

\section{Examples}

Let us first consider the following
simple PDF's characterized by the
parameters $T$,$T_1$ and $A$:

\begin{equation}
\label{C22} f_3^{(3)}(x_1,x_2,x_3) \propto
e^{-|\frac{x_1 -x_2}{T_1}|}
e^{-|\frac{x_1 - x_3}{T_1}|}
e^{-|\frac{x_2 - x_3}{T_1}|}),
\end{equation}

\begin{equation}
\label{C23} f_n^{(n)}(x_1,x_2,...,x_n) \propto
\prod_{i=1}^n e^{-\frac{x_i}{T}} (1 +
A\sum_{i<j\le n}
e^{-|\frac{x_i - x_j}{T_1}|}),
\end{equation}
The parameter $A$ in Eq.~(\ref{C23}) controls the
correlation strength; for $A = 0$ there are
no correlations among particles, so $C_l^{(x)}=0$.
Eq.~(\ref{C22}) corresponds to
two-particle correlations only, thus leading to $C_3^{(x)}=0$.

The correlator expectation values calculated according to
PDF's in Eqs. (\ref{C22})
and (\ref{C23}) by direct numerical integration and their
estimates obtained from simulated events are compared in
table \ref{tab-anal}.
\begin{table}
  \begin{center}
    \caption{The correlators corresponding to PDF's
    in Eqs.~(\ref{C22}) (the first two lines)
    and (\ref{C23}). The results of numerical
    integration are compared with the estimates obtained from
    $n_{\rm evt}$ simulated events.
    The allowed interval of the variable $x$
    was 0-150, the parameters $T=150$, $T_1=25$ and $A=0,1,2$.
    }
    \label{tab-anal}
    \vspace*{0.5cm}

\begin{tabular}{|c|c|c|c|}
\hline
$Correlator$ & $C_2^{(x)}$ &$ C_3^{(x)}$ &$ C_4^{(x)}$ \\
\hline
 $n=3$       &$1312.5$ &   $0.$    &           \\
$n_{\rm evt}=5\cdot 10^5$ & $1313.4 \pm 1.8$ &$61. \pm 108.$& \\
\hline
$n=3,A=0$          &$0.$ &   $0.$    &           \\
 $n_{\rm evt}=5\cdot 10^5$ &$ -0.3 \pm 1.4$&$21. \pm 70.$& \\
\hline
$ n = 3, A = 2$       &$229.0 $  & $1771.5$ &      \\
$n_{\rm evt}=5\cdot 10^5$ &$229.9 \pm 1.4$&$1779. \pm 110.$& \\
\hline
n = 4, A = 1     &$ 113.8$&$707. $&  $2913.$\\
$n_{\rm evt}=3\cdot 10^6$&$ 114.4 \pm 0.5$&$666. \pm 23.$
&  $3890.\pm 1836.$\\
\hline
\end{tabular}
\end{center}
\end{table}
One may see that the estimated values of the correlators agree
with their expectation values within the errors.
The relative errors rapidly increase with the order $l$ of the
correlator thus making quite uneasy its measurement
for $l > 4$.


As another example, we consider the energy correlators
estimated from the ``microcanonical''
ensemble of events simulated according to the
non-relativistic phase space (non-relativistic ideal gas of
particles) using Metropolis algorithm to redistribute
the particle energies via binary energy-conserving collisions.
The total
amount of energy distributed to $n_{\rm tot}$ particles is
$X_{\rm tot}=n_{\rm tot} \bar{x} $, where
$\bar{x}$ is the mean particle energy.
We have put $\bar{x} =100$ in arbitrary units.
Assuming that all $n_{\rm tot}$ particles are observed,
the mean particle energy does not
fluctuate ($\Delta\bar{x}=0$) and the correlators $C_l^{(x)}$
are then given
by Eqs.~(\ref{KK2})-(\ref{KK4}) with $n=n_{\rm tot}$.
Particularly, $C_2^{(x)}$ and $C_3^{(x)}$
vanish at large $n_{\rm tot}$ as $1/n_{\rm tot}$ and
$1/n_{\rm tot}{}^2$ respectively.
The results shown in table \ref{tab-micro} confirm this behavior.

\begin{table}
  \begin{center}
    \caption{Correlators $C_2^{(x)}$ and $C_3^{(x)}$
    estimated from the ``microcanonical'' ensemble of events
    with fixed mean particle energy $\bar{x} =100$ in arbitrary
    units,
    each consisting of $2\cdot 10^4$ events with fixed particle
    multiplicity $n_{\rm tot}$.
    All particles are assumed to be observed, i.e.
    $n=n_{\rm tot}$.
    }
    \label{tab-micro}
    \vspace*{0.5cm}

\begin{tabular}{|c|c|c|c|c|c|}
\hline
$n_{\rm tot}$ & 5 & 10 & 15 & 20 & 100\\
\hline
$C_2^{(x)}$ & $-1181 \pm 2$ & $-624 \pm 1$ & $-426 \pm 1$
& $-323 \pm 1$ & $-66.2 \pm 0.1$\\
\hline
$C_3^{(x)}$ & $49900 \pm 200$ & $14730 \pm 50$ & $6940 \pm 20$
& $4041 \pm 6$ & $174 \pm 1$\\
\hline
\end{tabular}
\end{center}
\end{table}

\begin{table}
  \begin{center}
    \caption{The same as in table \ref{tab-micro} for
    $n_{\rm tot}= 100$ and different numbers $\nu$ of
    selected particles.
    }
    \label{tab-micro1}
    \vspace*{0.5cm}
\begin{tabular}{|c|c|c|c|c|}
\hline
$\nu$ & 5 & 10 & 15 & 20\\
\hline
$C_2^{(x)}$ & $-75 \pm 8$ & $-63 \pm 3$ & $-68 \pm 2$ & $-66 \pm 2$\\
\hline
$C_3^{(x)}$ & $200 \pm 500$ & $300 \pm 100$ & $220 \pm 70$
& $170 \pm 50$\\
\hline
\end{tabular}
\end{center}
\end{table}
In table \ref{tab-micro1}, we show the same correlators as in
table \ref{tab-micro} for $n_{\rm tot}= 100$, but now calculated
for different numbers of selected particles $\nu<n_{\rm tot}$.
One may see that within the errors the correlator estimates are
$\nu$-independent. This confirms the conclusion at the end of
section \ref{Sec_cor-fluct} about the
compensation of the $\nu$-dependence of the quantities
$c_l^{(\nu,x)}$ by that of the $\bar{x}$ fluctuations.

Finally, we have estimated the pion transverse momentum correlators
$C_2^{(p_t)}$, $C_3^{(p_t)}$ and $C_4^{(p_t)}$
using $2 \cdot 10^3$ central (impact parameter $b = 0$ fm) events
of $Pb+Pb$ collisions at total c.m. energy $\sqrt{s} = 200$ AGeV
simulated in the parton string model (PSM) \cite{Am2001}.
In this phenomenological model, the soft and semihard parton
collisions initiate
the formation of color strings and the subsequent string breaking
leads to the production of stable (with respect to strong interaction)
hadrons and resonances that
are forced to decay.
We have divided the simulated events into $100$ subgroups
and selected $\nu = 1900$  pions of a given
charge in each event.
As one can see from table \ref{tab-PSM}, the correlators $C_3^{(p_t)}$
are consistent with zero within the errors while this is not the
case for the correlators $C_2^{(p_t)}$  and $C_4^{(p_t)}$.
The latter appear to be non-zero for any combination of pion
charges, being somewhat higher for neutral pions.
\begin{table}
  \begin{center}
    \caption{The estimates of the pion transverse momentum
    correlators obtained from $2 \cdot 10^3$ simulated central PSM events
     of $Pb+Pb$ collisions at total c.m. energy $\sqrt{s} = 200$ AGeV.
    }
    \label{tab-PSM}
    \vspace*{0.5cm}

\begin{tabular}{|c|c|c|c|}
\hline
Species & $C_2^{(p_t)}$ &$ C_3^{(p_t)}$ &$ C_4^{(p_t)}$ \\
&(GeV/c)$^2$&(GeV/c)$^3$&(GeV/c)$^4$\\
\hline
$\pi^+$ &$(44.0 \pm 6.6)\cdot 10^{-6}$  &$(-7.4 \pm 12.1)\cdot 10^{-8}$& $(8.7 \pm 3.3)\cdot 10^{-9}$\\
\hline
$\pi^-$ &$(44.6 \pm 6.0)\cdot 10^{-6}$&$(7.8 \pm 12.5)\cdot 10^{-8}$& $(9.5 \pm 4.6)\cdot 10^{-9}$\\
\hline
$\pi^0$ &$(71.2 \pm 7.6)\cdot 10^{-6}$&$ (3.0 \pm 2.7)\cdot 10^{-7}$& $(34.3 \pm  9.5)\cdot 10^{-9}$\\
\hline
$\pi^{\pm}$ &$(52.6 \pm 4.2)\cdot 10^{-6}$&$ (-1.5 \pm 7.1)\cdot 10^{-8}$& $(9.0 \pm 2.0)\cdot 10^{-9}$\\
\hline
$\pi^{\pm,0}$ &$(59.6 \pm 3.5)\cdot 10^{-6}$&$ (2.1 \pm 5.3)\cdot 10^{-8}$& $(11.2 \pm 1.6)\cdot 10^{-9}$\\
\hline
\end{tabular}
\end{center}
\end{table}
We have also selected $\pi^+\pi^-$ pairs only and estimated
$C_{+-}^{(p_t)} = (61.9 \pm 4.4)\cdot 10^{-6} (GeV/c)^2$.
Since the PSM takes into account only the resonances from the
lowest SU(3) multiplets that do not decay into pairs of
like-sign charged pions,
the small difference between the correlators
for pairs of like-sign and unlike-sign charged pions
indicates that the
non-zero PSM correlators are mainly of the non-resonance origin.
Also, since we consider the high multiplicity events, the
correlations due to energy-momentum conservation
are of minor importance in accordance with the discussion
in section \ref{Sec_cor-fluct}.
The probable source of the non-zero PSM correlators are thus
the semihard parton collisions that are becoming important with
the increasing energy and are known to lead to a noticeable
"non-statistical" $\bar{p}_t$ fluctuation in $Au+Au$
collisions at RHIC \cite{STAR2003}.


\section{Conclusion}

We have developed a fast procedure allowing one to calculate,
in a reasonable computer time,
the particle correlators of any order
and estimate their errors.
The corresponding C++ code is available on the request at
e-mail address amelin@sunhe.jinr.ru.
We have suggested the extension of this procedure for the
event-by-event approach as well.
We have shown a close relation between the correlators and
fluctuations of the observable event-mean values.
We have applied the proposed procedure to the events simulated
within various models and demonstrated the usefulness of the
two-, three- and four-particle correlators; the measurement of
the higher-order correlators is rather difficult as
it requires very high statistics.

\section*{Acknowledgements}
We are grateful to J. Manjavidze and S. Shimanskii
for useful discussions and - to
A. Sissakian and Yu. Panebratsev for
drawing our attention to this topic.
This work was supported by the Grant Agency of the Czech Republic
under contract 202/04/0793 and
Slovak Grant Agency for Sciences under contract 2/4099/24.

%
\newcounter{appnum}
\setcounter{appnum}{0}
\def\theappnum{\Alph{appnum}}

\def\makeappendix#1{%
\stepcounter{appnum}
\addcontentsline{toc}{section}
{\appendixname\space\theappnum #1}
\setcounter{equation}{0}%
\gdef\theequation{%
\theappnum.\arabic{equation}}%
\section*{\appendixname\ \theappnum #1}}
\def\appendixname{Appendix}
%
\makeappendix{}
To clarify the meaning of quantities $c_l^{(x)}$,
let us follow Ref. \cite{Sch} and
consider the powers of the equality
\begin{equation}
\label{ident1}
0=\sum_{j=1}^{n}(x_j-\bar{x})\equiv
\sum_{j=1}^{n}\Delta_j .
\end{equation}
Thus, the second, third and fourth powers of
Eq.~(\ref{ident1}) yield
\begin{equation}
\label{ident2}
0=\sum_{j\ne k}\Delta_j\Delta_k +\sum_{j}\Delta_j{}^2,
\end{equation}
\begin{equation}
\label{ident3}
0=\sum_{j\ne k\ne l}\Delta_j\Delta_k\Delta_l +
{3\choose 2}\sum_{j\ne k}\Delta_j\Delta_k{}^2 +
\sum_{j}\Delta_j{}^3,
\end{equation}
$$
0=\sum_{j\ne k\ne l\ne m}\Delta_j\Delta_k\Delta_l\Delta_m +
{4\choose 2}\sum_{j\ne k\ne l}\Delta_j\Delta_k\Delta_l{}^2 +
$$
\begin{equation}
\label{ident4}
\frac{1}{2!}{4\choose 2}{4-2\choose 2}\sum_{j\ne k}\Delta_j{}^2\Delta_k{}^2 +
{4\choose 3}\sum_{j\ne k}\Delta_j\Delta_k{}^3 +
\sum_{j}\Delta_j{}^4.
\end{equation}
Using further the relations
\begin{equation}
\label{ident3a}
\sum_{j\ne k}\Delta_j\Delta_k{}^\lambda =
-\sum_{k}\Delta_k{}^{\lambda+1},
\end{equation}
$$
\sum_{j\ne k\ne l}\Delta_j\Delta_k\Delta_l{}^2 =
\left(\sum_{j\ne k}\Delta_j\Delta_k\right)\sum_l\Delta_l{}^2-
2\sum_{j\ne k}\Delta_j\Delta_k{}^3
$$
\begin{equation}
\label{ident4a}
= -\left(\sum_l\Delta_l{}^2\right)^2 +
2\sum_{l}\Delta_l{}^4,
\end{equation}
and
\begin{equation}
\label{ident4b}
\sum_{j\ne k}\Delta_j{}^2\Delta_k{}^2
= \left(\sum_l\Delta_l{}^2\right)^2 -
\sum_{l}\Delta_l{}^4,
\end{equation}
one can express the multiple sums in Eq.~(\ref{C25'''})
through the sums of the powers of $\Delta_j{}$
related to the estimates of the central moments
$m_\lambda^{(x)}$ in Eq.~(\ref{schc5}).
Rewriting Eq.~(\ref{C25'''}) in the form
\begin{equation}
\label{C25'''b} c^{(i,x)}_l =
\frac{1}{n(n-1)\dots (n-l+1)}
\sum_{i_1\ne i_2\ne\dots\ne i_l}
\Delta_{i_1}\Delta_{i_2}\dots\Delta_{i_l},
\end{equation}
one then proves Eqs.~(\ref{KK2})-(\ref{KK4}) relating
quantities $c_l^{(x)}$ with central moments
$m_\lambda^{(x)}$.



\begin{thebibliography}{99}
\bibitem{Voloshin2002}
C. Pruneau, S. Gavin and S. Voloshin, Phys. Rev. C 66 (2002) 044904.
\bibitem{STAR2003}
J. Adams et al. (STAR), nucl-ex/0308033.

\bibitem{QM04} Quark Matter 2004 Conference Proceedings,
J.Phys. G30 (2004) Suppl.

\bibitem{MS2001} J. Manjavidze and A. Sissakian, Phys. Rep. {\bf 346} (2001) 1.

\bibitem{Am2001}
N. Amelin, N. Armesto, C. Pajares and D. Sousa,
Eur. Phys. J. {\bf C22}
(2001) 149.

\bibitem{Sch} A.V. Stadnik, N.I. Chernov and S.S. Shimanskii,
JINR Report P11-2003-143, 2003.




\end{thebibliography}
\end{document}